# A NOVEL APPROACH TOWARDS COST EFFECTIVE REGION-BASED GROUP KEY AGREEMENT PROTOCOL FOR PEER -TO - PEER INFORMATION SHARING IN MOBILE AD HOC NETWORKS

K. Kumar[1], J.Nafeesa Begum[2] and Dr.V. Sumathy[3]

1. Research Scholar & Asst .Professor in CSE, Government College of Engg, Bargur- 635104, Tamil Nadu,India
*pkk_kumar@yahoo.com*
2. Research Scholar & Asst .Professor (Sr) in CSE, Government College of Engg, Bargur- 635104, Tamil Nadu, India
*nafeesa_jeddy@yahoo.com*
3. Associate Professor in ECE,Government College of Technology,Coimbatore, Tamil Nadu,India
*sumi_gct2001@yahoo.co.in*

## ABSTRACT

*Peer-to-peer systems have gained a lot of attention as information sharing systems for the wide-spread exchange of resources and voluminous information that is easily accessible among thousands of users. However, current peer-to-peer information sharing systems work mostly on wired networks. With the growing number of communication-equipped mobile devices that can self-organize into infrastructure-less communication platform, namely mobile ad hoc networks (MANETs), peer-to-peer information sharing over MANETs becomes a promising research area. In this paper, we propose a Region-Based structure that enables efficient and secure peer-to-peer information sharing over MANETs. The implementation shows that the proposed scheme is Secure, scalable, efficient, and adaptive to node mobility and provides Reliable information sharing.*

## 1. INTRODUCTION

A peer –to – peer (P2P) system is a self organizing system of equal, autonomous entities which aims for the shared usage of distributed resources in networked environment avoiding central services. The P2P networks are a powerful architecture for sharing of resources and voluminous information among thousands of users. An important issue in P2P system is searching for resources (e.g., data, files and services) available at one or more of the numerous host nodes. The distributed nature of P2P systems can be an advantage over client-server architectures due to the following reasons.  First, they tend to be more fault-tolerant as there is no single point – of – failure. Another important reason is processing of information, network traffic and data storage can be balanced over all peers, which enables the network to scale well with the number of peers.

Group Key Agreement (GKA) protocols [1,3], which enable the participants to agree on a common secret value, based on each participant's public contribution, seem to provide a good solution.  They don't require the presence of a central authority. Also, when the group composition changes, Group Controller can employ supplementary key agreement protocols to get a new group key.
Elliptic Curve Cryptography (ECC)[2,7] is a public key cryptosystem based on elliptic curves. The attraction of ECC is that it appears to offer equal security for a far smaller key size, thereby reducing processing overhead.

In this paper, we propose a reliable and secure Region-Based Key Agreement Protocol for peer-to-peer information sharing. Here, we break a group into region-based subgroups with leaders in subgroups communicating with each other to agree on a group key in response to membership change. In addition to showing that the forward and backward secrecy requirements are satisfied, we identify optimal settings of our protocol to minimize the overall communication and computation costs due to group key management.

The contribution of this work includes:

1. In this paper, we propose a new efficient method for solving the group key management problem for effective P2P information sharing over mobile ad-hoc network. This protocol provides efficient, scalable, reliable and secure P2P information sharing over ad-hoc network.

2. We introduce the idea of subgroup and subgroup key and we uniquely link all the subgroups into a tree structure to form an outer group and outer group key. This design eliminates the centralized key server. Instead, all hosts work in a peer-to-peer fashion to agree on a group key. Here we propose a new protocol ECRBGKA (Elliptic Curve Region-Based Key Agreement) for ad hoc networks. It is a combination of GECDH & TGECDH protocol so as to keep the group key secure among the group members in P2P group communication with a shorter key length and same security level as that of other cryptosystems.

3. We design and implement ECRBGKA protocol in peer-to-peer information sharing using Java and the extensive experiments show that the secure P2P information sharing in ad-hoc networks is achievable in addition to the performance issues like memory cost, communication cost and computation cost.

The rest of the paper is as follows, Section 2 presents the Proposed Schemes. Section 3 describes the message encryption and decryption using ECC. Section 4 describes the Experimental Results and Discussion. Section 5 describes the performance analysis. and finally Section 6 concludes the paper.

## 2. PROPOSED SCHEME

### 2.1. System Model

In this section we first provide an overview of our secure information sharing system, including the security model.

#### 2.1.1. Overview of Region-Based Group Key Agreement Protocol:

The goal of this paper is to propose a communication and computation efficient group key establishment protocol in ad-hoc network. The idea is to divide the multicast group into several subgroups, let each subgroup has its subgroup key shared by all members of the subgroup. Each Subgroup has subgroup controller node and a Gateway node, in which Subgroup controller node is the controller of subgroup and each Gateway node contributes a partial key to agree with a common outer group key among the subgroups. The last gateway member joining the outer group acts as outer group controller.

For example, in Figure.1, all member nodes are divided into number of subgroups and all subgroups are linked in a tree structure as shown in Figure.2.

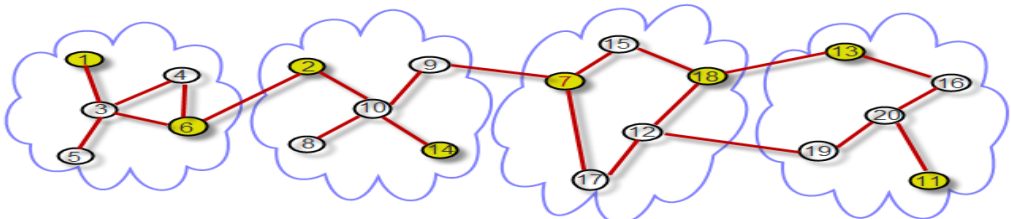

Figure.1: Members of group are divided into subgroups

International Journal of peer-to-peer networks (IJP2P) Vol.1, No.1, September 2010

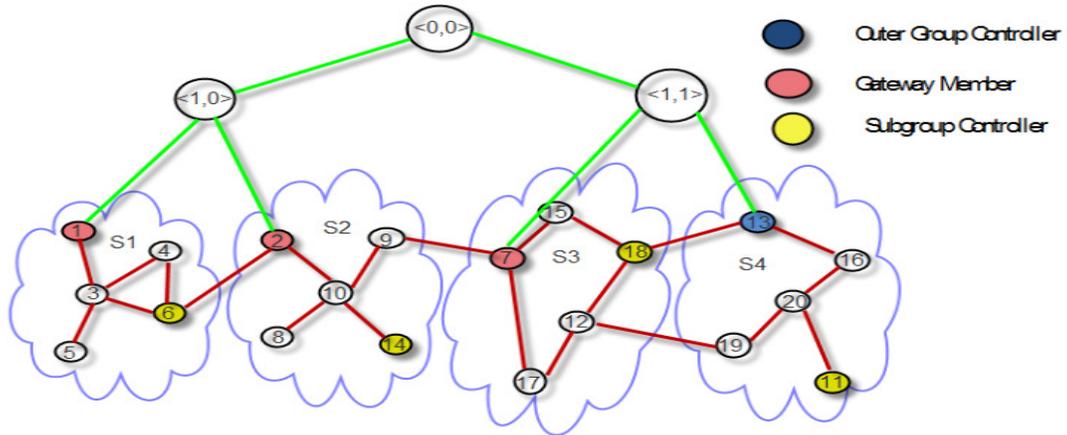

Figure.2: Subgroups link in a Tree Structure

The layout of the network is as shown in below Figure.3.

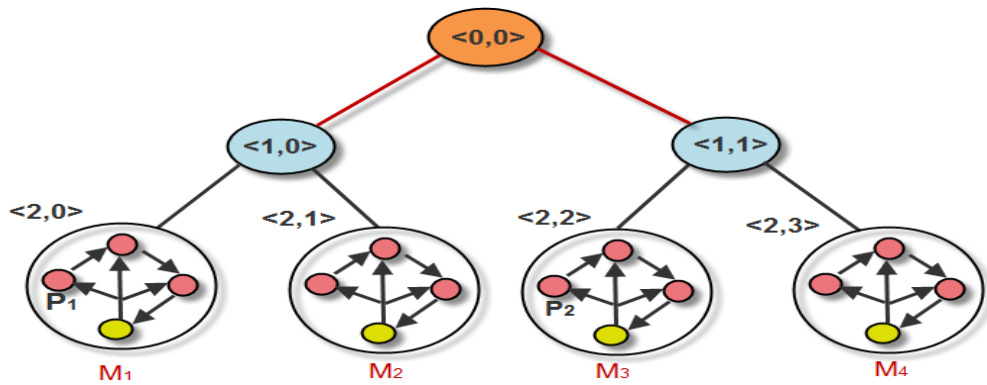

Figure.3. Region based Group Key Agreement

One of the members in the subgroup is subgroup controller. The last member joining the sub group acts as a subgroup controller. Each outer group is headed by the outer group controller. In each group, the member with high processing power, memory, and Battery power acts as a gateway member. Outer Group messages are broadcast through the outer group and secured by the outer group key while subgroup messages are broadcast within the subgroup and secured by subgroup key.

Let N be the total number of group members, and M be the number of the subgroups in each subgroup, then there will be N/M subgroups, assuming that each subgroup has the same number of members.

Assume that there are a total of N members in Secure Group Communication. After sub grouping process (Algorithm 1), there are S subgroups $M_1, M_2 \ldots M_s$ with $n_1, n_2 \ldots n_s$ members.

---

**Algorithm. 1. Region-Based Key Agreement protocol**

1. The Subgroup Formation
   The number of members in each subgroup is
   $N / S < 100.$
   Where, N – is the group size. and  S – is the number of subgroups.

2. The Contributory Key Agreement protocol is implemented among the group members. It consists of three stages.
   a. To find the Subgroup Controller for each subgroups.
   b. GECDH protocol is used to generate one common key for each subgroup headed by the subgroup controller(i.e Sub group Key (KR)),which performs encryption and decryption of sub group level messages broadcast to all subgroup members..
   c. Each subgroup gateway member contributes partial keys to generate a one common backbone key (i.e Outer group Key (KG)) ,which is used to encrypt and decrypt the messages broadcast among subgroup controllers, headed by the Outer Group Controller using TGECDH protocol.

3. Each Group Controller (Sub /Outer) distributes the computed public key to all its members. Each member performs rekeying to get the respected group key.

---

A Regional key KR is used for communication between a subgroup controller and the members in the same region. The Regional key KR is rekeyed whenever there is a membership change event occurs due to member join / leave or member failure. The Outer Group key KG is rekeyed whenever there is a join / leave of Outer group controllers or Gateway member failure to preserve secrecy.

The members within a subgroup use Group Elliptic Curve Diffie-Hellman Contributory Key Agreement (GECDH). Each member within a subgroup contributes his share in arriving at the subgroup key. Whenever membership changes occur, the subgroup controller or previous member initiates the rekeying operation.

The gateway member initiates communication with the neighboring member belonging to another subgroup and mutually agree on a key using Tree-Based Group Elliptic Curve Diffie-Hellman contributory Key Agreement (TGECDH) protocol for inter subgroup communication between the two subgroups. Any member belonging to one subgroup can communicate with any other member in another subgroup through this member as the intermediary. In this way adjacent subgroups agree on outer group key. Whenever membership changes occur, the outer group controller or previous outer group controller initiates the rekeying operation.

Here, we prefer the subgroup key to be different from the key for backbone. This difference adds more freedom of managing the dynamic group membership. In addition, using this approach can potentially save the communication and computational cost.

An information sharing application should allow end users to:
- Search for information.
- Make their information available.
- Download information.

In order to secure this type of information sharing application we need to provide confidentiality and integrity of all communication and to enforce access control of resources i.e. communication channels and shared information.

Figure 3. shows the interactions between peers. A peer, $P_1$, searches for information by sending a *query* message to the subgroup. The subgroup controller sends the query message to outer group. Peers subgroup that do not have content that matches the query, do not respond. Peer $P_2$ has content that matches the query, so it sends a *query response* message. $P_1$, can then request to download content from $P_2$ by sending a *transfer request* message. $P_2$ answers this request with the content

**2.2 .Network Dynamics**

The network is dynamic in nature. Many members may join or leave the group. In such case, a group key management system should ensure that backward and forward secrecy is preserved.

### 2.2.1. Member Join

When a new member joins, it initiates communication with the subgroup controller. After initialization, the subgroup controller changes its contribution and sends public key to this new member. The new member receives the public key and acts as a group controller by initiating the rekeying operations for generating a new key for the subgroup. The rekeying operation is as follows.

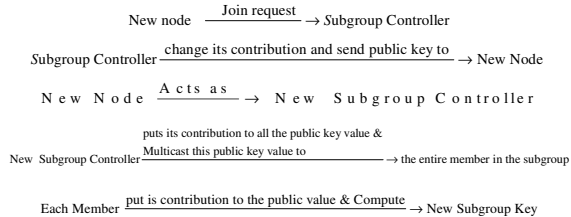

$$\text{New node} \xrightarrow{\text{Join request}} \text{Subgroup Controller}$$
$$\text{Subgroup Controller} \xrightarrow{\text{change its contribution and send public key to}} \text{New Node}$$
$$\text{New Node} \xrightarrow{\text{Acts as}} \text{New Subgroup Controller}$$
$$\text{New Subgroup Controller} \xrightarrow{\substack{\text{puts its contribution to all the public key value \&}\\ \text{Multicast this public key value to}}} \text{the entire member in the subgroup}$$
$$\text{Each Member} \xrightarrow{\text{put is contribution to the public value \& Compute}} \text{New Subgroup Key}$$

### 2.2.2. Member Leave:
### 1. When a Subgroup member Leaves

When a member leaves the Subgroup Key of the subgroup to which it belongs must be changed to preserve the forward secrecy. The leaving member informs the subgroup controller. The subgroup controller changes its private key value, computes the public value and broadcasts the public value to all the remaining members. Each member performs rekeying by putting its contribution to public value and computes the new Subgroup Key. The rekeying operation is as follows.

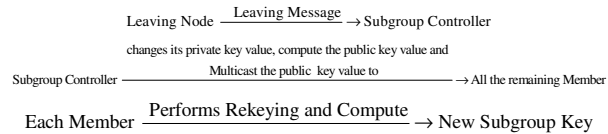

$$\text{Leaving Node} \xrightarrow{\text{Leaving Message}} \text{Subgroup Controller}$$
$$\text{Subgroup Controller} \xrightarrow{\substack{\text{changes its private key value, compute the public key value and}\\ \text{Multicast the public key value to}}} \text{All the remaining Member}$$
$$\text{Each Member} \xrightarrow{\text{Performs Rekeying and Compute}} \text{New Subgroup Key}$$

### 2. When a Subgroup Controller Leaves:

When the Subgroup Controller leaves, the Subgroup key used for communication among the subgroup controller needs to be changed. This Subgroup Controller informs the previous Subgroup Controller about its desire to leave the subgroup which initiates the rekeying procedure. The previous subgroup controller now acts as a Subgroup controller. This Subgroup controller changes its private contribution value and computes all the public key values and broadcasts to all the remaining members of the group. All subgroup members perform the rekeying operation and compute the new subgroup key. The rekeying operation is as follows.

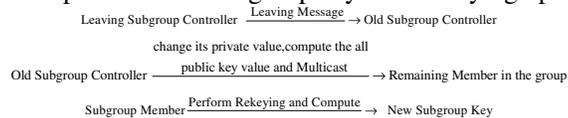

$$\text{Leaving Subgroup Controller} \xrightarrow{\text{Leaving Message}} \text{Old Subgroup Controller}$$
$$\text{Old Subgroup Controller} \xrightarrow{\substack{\text{change its private value, compute the all}\\ \text{public key value and Multicast}}} \text{Remaining Member in the group}$$
$$\text{Subgroup Member} \xrightarrow{\text{Perform Rekeying and Compute}} \text{New Subgroup Key}$$

### 3. When an Outer Group Controller Leaves:

When an Outer group Controller leaves, the Outer group key used for communication among the Outer group need to be changed. This Outer group Controller informs the previous Outer group Controller about its desire to leave the Outer group which initiates the rekeying procedure. The previous Outer Group controller now becomes the New Outer group controller. This Outer group controller changes its private contribution value and computes the public key value and broadcasts it to the entire remaining member in the group. All Outer group members perform the rekeying operation and compute the new Outer group key. The rekeying operation is as follows.

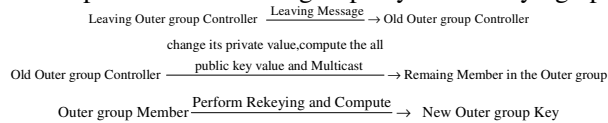

$$\text{Leaving Outer group Controller} \xrightarrow{\text{Leaving Message}} \text{Old Outer group Controller}$$
$$\text{Old Outer group Controller} \xrightarrow{\substack{\text{change its private value, compute the all}\\ \text{public key value and Multicast}}} \text{Remaing Member in the Outer group}$$
$$\text{Outer group Member} \xrightarrow{\text{Perform Rekeying and Compute}} \text{New Outer group Key}$$

### 4. When a Gateway member leaves

When a gateway member leaves the subgroup, it delegates the role of the gateway to the adjacent member having high processing power, memory, and Battery power and acts as a new gateway member. Whenever the gateway member leaves, all the two keys should be changed. These are
   i.   Outer group key among the subgroup.
   ii.  Subgroup key within the subgroup.

In this case, the subgroup controller and outer group controller perform the rekeying operation. The Controller leaves the member and a new gateway member is selected in the subgroup, performs rekeying in the subgroup. After that, it joins in the outer group. The procedure is same as joining the member in the outer group.

### 2.3. Communication Protocol:

The members within the subgroup have communication using subgroup key (KR). The communication among the subgroup members takes place through the gateway member using the Outer Group Key (KG).

### 2.3.1. Communication within the Subgroup:
   **1. Sender query search within the Subgroup**

The sender member encrypts the message with the subgroup key (KR) and multicasts it to all member in the subgroup. The subgroup members receive the encrypted message, perform the decryption using the subgroup key (KR) and gets the original message. The communication operation is as follows.

$$\text{Source Member} \xrightarrow{E_{KR}[\text{Message}] \ \& \ \text{Multicast}} \text{Destination Member}$$

$$\text{Destination Member} \xrightarrow{D_{KR}[E_{KR}[\text{Message}]]} \text{Query Message}$$

  **2. Response for Sender Query**

The Destination member encrypts the response message with the Subgroup Key (KR) and multicast it to all member in the subgroup. The source member receives the encrypted message, performs the decryption using the subgroup key and gets the response message. The communication operation is as follows.

$$\text{Destination Member} \xrightarrow{E_{KR}[\text{Message}] \ \& \ \text{Multicast}} \text{Source Member}$$

$$\text{Source Member} \xrightarrow{D_{KR}[E_{KR}[\text{Message}]]} \text{Response Message}$$

### 2.3.2. Communication among the Subgroup:
   **1. Sender query search among the Subgroup**

The sender member encrypts the message with the subgroup key (KR) and multicasts it to all members in the subgroup. One of the members in the subgroup acts as a gate way member. This gateway member decrypts the message with subgroup key and encrypts with the outer group key (KG) and multicast to the entire gateway members in the group. The destination gateway member first decrypts the message with outer group key. Then encrypts with subgroup key and multicasts it to all members in the subgroup. Each member in the subgroup receives the encrypted message and performs decryption using subgroup key and gets the original message. In this way the region-based group key agreement protocol performs the communication. The communication operation is as follows.

$$\text{Source Member} \xrightarrow{E_{KR}[\text{Message}] \ \& \ \text{Multicast}} \text{Gateway Member}$$

$$\text{Gateway Member} \xrightarrow{D_{KR}[E_{KR}[\text{Message}]]} \text{Query Message}$$

$$\text{Gateway Member} \xrightarrow{E_{KG}[\text{Message}] \ \& \ \text{Multicast}} \text{Gateway Member [ Among Subgroup]}$$

$$\text{Gateway Member} \xrightarrow{D_{KG}[E_{KG}[\text{Message}]]} \text{Query Message}$$

$$\text{Gateway Member} \xrightarrow{E_{KR}[\text{Message}] \ \& \ \text{Multicast}} \text{Destination Member}$$

$$\text{Destination Member} \xrightarrow{D_{KR}[E_{KR}[\text{Message}]]} \text{Query Message}$$

  **2. Response for Sender Query**

The receiver member encrypts the response message with its subgroup key (KR) and multicasts it to all members in the subgroup. The gateway member decrypts the response message with the subgroup key and encrypts with the Outer group key (KG) and multicast to the entire gateway members in the group. The destination gateway member first decrypts the message with the Outer group key. Then encrypts with subgroup key and multicast it to all members in the subgroup. Sender member in the subgroup receives the encrypted message and performs decryption using subgroup key and gets the response message. The communication operation is as follows.

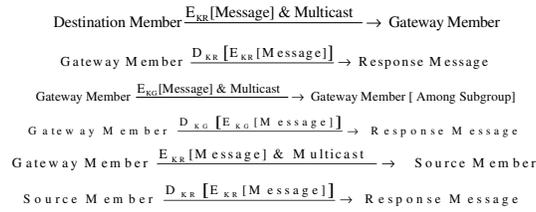

$$\text{Destination Member} \xrightarrow{E_{KR}[\text{Message}] \& \text{Multicast}} \text{Gateway Member}$$
$$\text{Gateway Member} \xrightarrow{D_{KR}[E_{KR}[\text{Message}]]} \text{Response Message}$$
$$\text{Gateway Member} \xrightarrow{E_{KG}[\text{Message}] \& \text{Multicast}} \text{Gateway Member [ Among Subgroup]}$$
$$\text{Gateway Member} \xrightarrow{D_{KG}[E_{KG}[\text{Message}]]} \text{Response Message}$$
$$\text{Gateway Member} \xrightarrow{E_{KR}[\text{Message}] \& \text{Multicast}} \text{Source Member}$$
$$\text{Source Member} \xrightarrow{D_{KR}[E_{KR}[\text{Message}]]} \text{Response Message}$$

## 2.4. Applying Elliptic Curve based Diffie-Hellman Key Exchange

### 2.4.1. Member Join

User A and user B are going to exchange their keys (figure.4): Take p=211, Ep(0,-4), which is equivalent to the curve $y^2=x^3 - 4$ and G = (2,2). A's private key is nA = 47568, so A's public key PA =47568(2,2)=(206,121), B's private key is nB = 13525,so B's public key PB =13525(2,2)=(29,139). The group key is computed (Fig.[ ].) as User A sends its public key (206,121) to user B, then user B computes their Subgroup key as nB (A's Public key ) = 13525(206,121) = (**155,115**). User B sends its public key (29,139) to User A, then User A compute their Subgroup key as nA(B's Public key)= 47568(29,139) = (**120,180**) .The implementation is shown in figure 7.

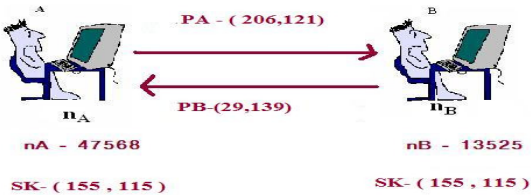

Figure.4.User-A & User –B Join the Group.

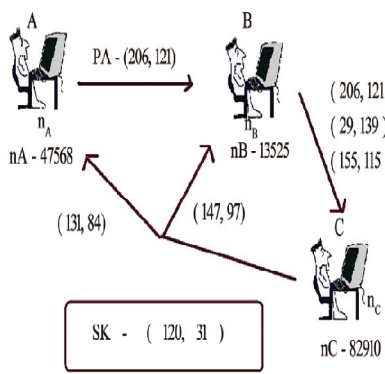
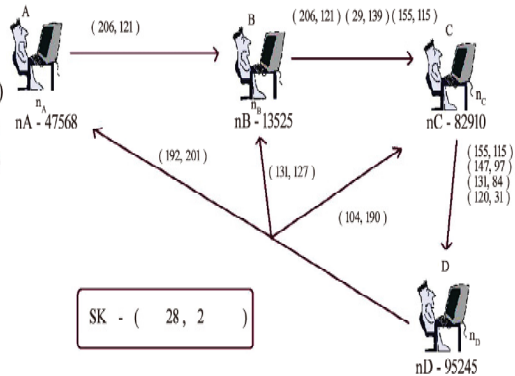

Figure .5. User- C Join in the Group.    Figure.6. User-D Join in the Group.

When User C is going to join in the group, C's private key becomes nC= 82910. Now, User C becomes a Subgroup Controller. Then, the key updating process will begin as follows: The previous Subgroup Controller User B sends the intermediate key as (B's Public key $ A's Public Key $ Group key of A&B)= ((29,139) $ (206,121) $ (155,115)) User C separates the intermediate key as B's Public key, A's Public Key and Group key of A&B=(29,139) , (206,121) and (155,115).Then, User C generates the new Subgroup key as nC (Subgroup key of A&B)= 82910(155,115) = (**120,31**). Then, User C broadcasts the intermediate key to User A and User B. That intermediate key is ((Public key of B & C) $ (Public key of A & C)) = ((131,84) $(147, 97)). Now, User B extracts the value of public key of A & C from the value sent by User

C. Then User B compute the new Subgroup key as follows: nB (Public key of A&C)= 13525(147,97)= (**120,31**)**.** Similarly, User A extracts the value of public key of B & C from intermediate key, sent by User C. Then User A compute the new Subgroup key as follows: nA (public key of B&C) = 47568(131,84) = (**120,31**). Therefore, New Subgroup Key of A, B and C = **(120, 31)** as shown in the figure.5.The same procedure is followed when User D joins as shown in the Fig.6.The implementation is shown in figure 8 and 9.

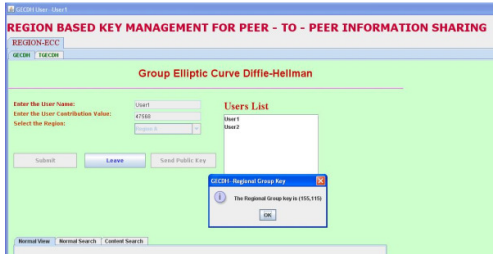

Figure 7. User 1 and User2 join the Subgroup Using GECDH

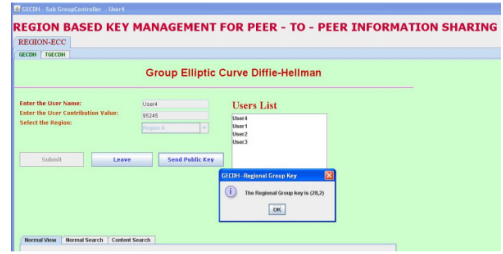

Figure 9. User4 join the Subgroup

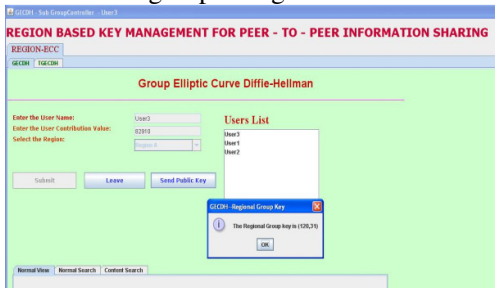

Figure 8. User 3 Join the Subgroup

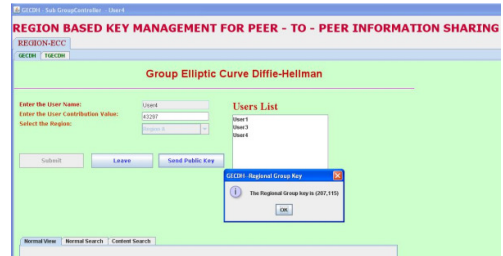

Figure 10. User2 leave from the Subgroup

### 2.4.2. Member Leave

When a user leaves (Fig.11.) from the Subgroup, then the Subgroup controller changes its private key. After that, it broadcasts its new public key value to all users in the Subgroup. Then, new Subgroup key will be generated. Let us consider, User B is going to leave, then the Subgroup Controller D changes its private key nD' =43297 ,so public key of User A & User C =(198,139)$(136,11). Then the new Subgroup Key generated is = 43297(198,139) = **(207,115).** Then, User A & User C computes the new Subgroup Key by using new public key. Therefore, the new Subgroup Key is **(207,115).** The implementation is shown in figure.10.

### 2.4.3. Group Controller Leave

When a Subgroup controller leaves (Fig.12.) from the group, then the previous Subgroup controller changes its private key. After that, it broadcasts its new public key value to all users in the group. Then, new Subgroup key will be generated. Let us consider, Subgroup Controller User D going to leave, then the previous Subgroup controller User C act as Subgroup Controller and changes its private key nC' = 52898 , and computes the public key of B&C $ A&C = (16,111)$(181,2). Then the new Subgroup Key generated is = 52898(21,103) = (**198,139**). Then, User A & User B compute the new Subgroup Key by using new public key. Therefore, the new Subgroup Key is **(198,139).** The implementation is shown in figure.13.

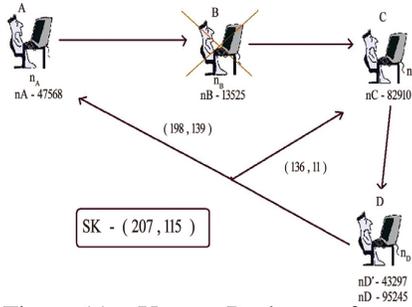

Figure.11. User –B leave from the Group.

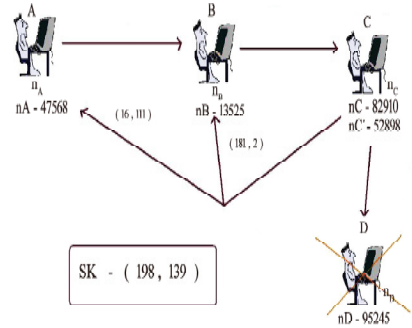

Figure.12. Group Controller Leave from the group.

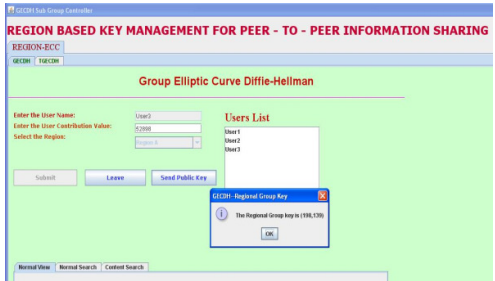

Figure 13. Subgroup Controller Leave from the Subgroup

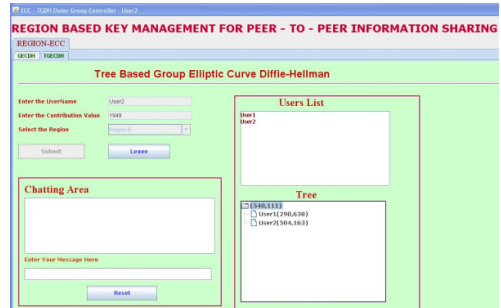

Figure 14. User 1 and User 2 Join the Outer group using TECGDH

### 2.5. Tree-based Group Elliptic Curve Diffie-Hellman Protocol

The proposed protocol (Fig.15.), Tree-based group Elliptic Curve Diffie-Hellman (TGECDH), is a variant of TGDH based on ECDLP. In TGECDH, a binary tree is used to organize group members. The nodes are denoted as $<l, v>$, where $0 \leq v \leq 2^l - 1$ since each level $l$ hosts at most $2^l$ nodes and $<l,v>$ means the v-th node at the l-th level on the key tree. Each node $<l, v>$ is associated with the key $K_{<l,v>}$ and the blinded key $BK_{<l,v>} = F(K_{<l,v>})$ where the function F() is scalar multiplication of elliptic curve points in prime field. Assuming a leaf node $<l, v>$ hosts the member $M_i$, the node $<l, v>$ has $M_i$'s session random key $K_{<l,v>}$. Furthermore, the member $M_i$ at node $<l, v>$ knows every key in the key-path from $<l, v>$ to $<0, 0>$. Every key $K_{<l,v>}$ is computed recursively as follows:

$$K_{<l,v>} = K_{<l+1,2v>} BK_{<l+1,2v+1>} \mod p$$
$$= K_{<l+1,2v+1>} BK_{<l+1,2v>} \mod p$$
$$= K_{<l+1,2v>} K_{<l+1,2v+1>} G \mod p$$
$$= F(K_{<l+1,2v>} K_{<l+1,2v+1>}) \quad (1)$$

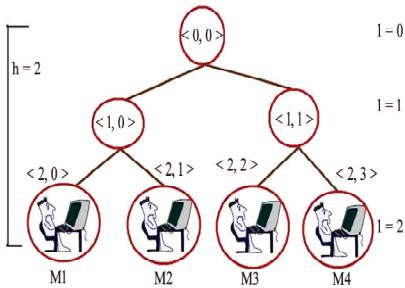

Figure.15. Key Tree.

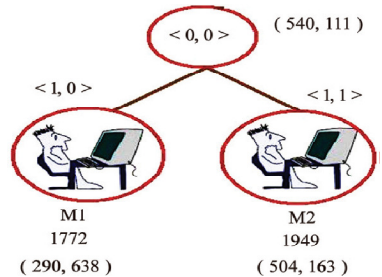

Figure.16. User $M_1$ & $M_2$ Join the Group

It is not necessary for the blind key $BK_{<l,v>}$ of each node to be reversible. Thus, simply use the x-coordinate of $K_{<l,v>}$ as the blind key. The group session key can be derived from $K_{<0,0>}$. Each time when there is member join/leave, the outer group controller node calculates the group session key first and then broadcasts the new blind keys to the entire group and finally the remaining group members can generate the group session key.

### 2.5.1. When node $M_1$ & $M_2$ Join the group.

User $M_1$ and User $M_2$ are going to exchange their keys: Take **p=751**, **Ep(1,188)**, which is equivalent to the curve $y^2=x^3+x+188$ and **G = (0,376)**. User $M_1$'s private key is **1772**, so $M_1$'s public key is **(290,638)**. User $M_2$'s private key is **1949**, so $M_2$'s public key is (504,163). The Outer Group key is computed (Figure.16) as User $M_1$ sends its public key **(290,638)** to user $M_2$, the User $M_2$ computes their group key as PV(0,0) = $X_{co}$ (PV(1,0) *PB(1,1)) and PB(0,0) = PV(0,0)*G =**(540,111)**. Similarly, User $M_2$ sends its public key **(504,163)** to user $M_1$, and then the user $M_1$ computes their group key as **(540,111)**. Here, Outer Group controller is User $M_2$. The implementation is shown in figure.14.

### 2.5.2. When 3$^{rd}$ node Join

When User $M_3$ joins the group, the old Outer group controller $M_2$ changes its private key value from **1949** to **2835** and passes the public key value and tree to User $M_3$. Now, $M_3$ becomes new Outer group controller. Then, $M_3$ generates the public key **(623, 52)** from its private key as **14755** and computes the Outer group key as **(664,736)** shown in Figure.17. $M_3$ sends Tree and public key to all users. Now, user $M_1$ and $M_2$ compute their group key. The same procedure is followed by joining the User $M_4$ as shown in Fig.18. The implementation is shown in figure 19 and 20.

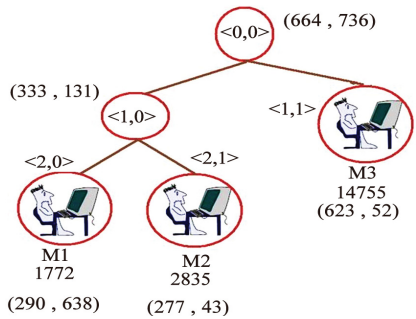 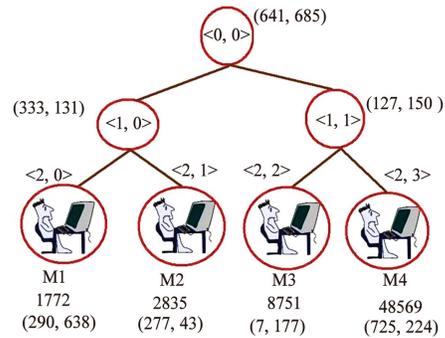

Figure.17. User $M_3$ Join the Group        Figure.18. User $M_4$ Join the group

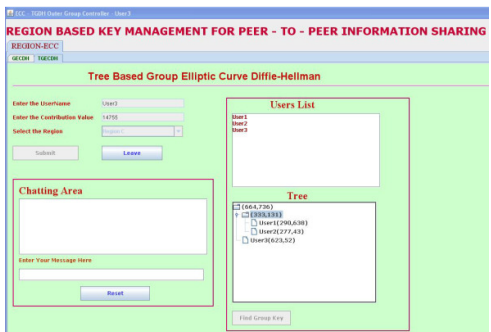 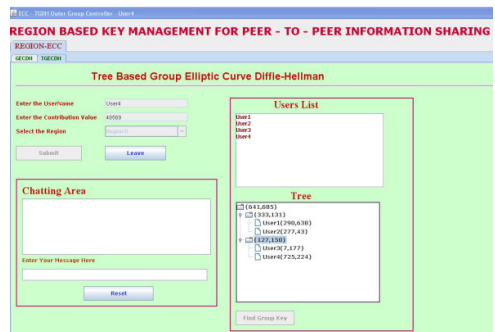

Figure 19. User 3 joins the Outer group     Figure 20. User 4 Joins the Outer group

### 2.5.3. Leave Protocol

There are two types of leave. 1. Gateway Member Leave and 2.Outer Group Controller Leave

## 1. Gateway Member Leave

When user $M_3$ leaves (Figure.21) the Outer group, then the Outer Group controller changes its private key **48569** to **98418** and outer group key is recalculated as **(428,686)**. After that, it broadcasts its Tree and public key value to all users in the Outer group. Then, the new Outer group key will be generated by the remaining users. The implementation is shown in figure.23.

## 2. When an Outer Group Controller Leaves

When an Outer Group Controller Leaves (Figure.22) from the group, then its sibling act as a New Outer Group Controller and changes its private key value 8751 to 19478 and recalculates the group key as (**681,475**). After that, it broadcast its Tree and public key value to all users in the Outer group. Then, the new Outer group key will be generated by the remaining users. The implementation is shown in figure 24.

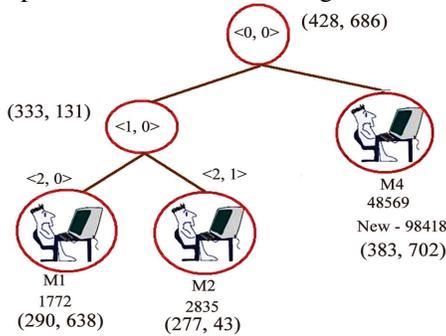
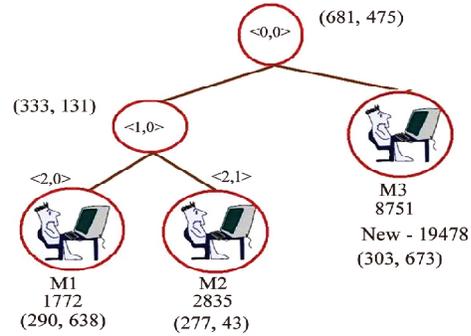

Figure.21. User $M_3$ Leave from the Outer Group

Figure.22. Outer Group Controller Leave from the Outer Group

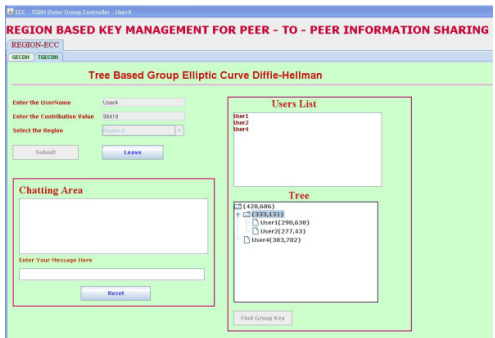
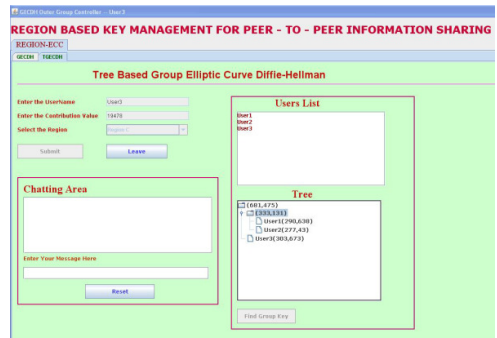

Figure 23. User 3 leave from the Outer group

Figure 24. Outer group controller leave from the Outer group

## 3. MESSAGE ENCRYPTION AND DECRYPTION

Elliptic curve cryptography can be used to encrypt plaintext message, M, into ciphertexts. The plaintext message M is encoded into a message point $P_M$ from the finite set of points in the elliptic group, $E_p(a,b)$. For example, User send the information to other user, first selects a random integer $k_A < n$ and computes $k_A G$ and $k_A S_K$. To encrypt the message $P_M$ to compute the ciphertext pair of points $P_c$:

$$P_C = [(k_A\ G), (P_M + k_A\ S_K)] \qquad (2)$$

After receiving the ciphertext pair of points, $P_C$, the receiver multiplies the first point, $(k_A\ G)$ with its shared secret key, $S_K$ and then adds the result to the second point in the ciphertext pair of points, $(P_M + k_A S_K)$:

$$(P_M + k_A S_K) - [S_K (k_A G)] = P_M + k_A S_K - k_A S_K = P_M \quad (3)$$

This is the plaintext point, corresponding to the plaintext message M.

Similarly, the gateway member received the message, first perform the decryption with subgroup key and then encrypt with outer group key KG. Other gateway members first decrypt the message with outer group key KG and then encrypt with its subgroup key SK.

Consider the example for message encryption and decryption.

### 3.1. Message Encryption:

Message             :    **dh1.png**
Random Number $K_A$  :    **75**
Cipher Text Pc      =    ( $K_A G$, $PM+K_A S_K$ )
$K_A G$  = 75 G = 75 mod 241  **=(32,108)**

**PM**
**d→100**  (93,77)
**h→104**  (163,50)
**1→49**   (207,96)
**. →46**  (83,124)
**p→112**  (58,12)
**n→110**  (164,197)
**g→103**  (67,154)

$S_k$ = (155,115) = 229 G
$K_A S_K$ = 75*229 G = 64 G

**PM+$K_A S_K$**

| | | | |
|---|---|---|---|
| d | → (100+64) mod 241 | = | (16,100) |
| h | → (104+64) mod 241 | = | (72,197) |
| 1 | → (49+64) mod 241  | = | (133,163) |
| . | → (46+64) mod 241  | = | (164,197) |
| p | → (112+64) mod 241 | = | (12,205) |
| n | → (110+64) mod 241 | = | (131,84) |
| g | → (103+64) mod 241 | = | (167,181) |

Therefore,

**$P_c$ = (32,108): (16,100):(72,197):(133,163):(164,197):(12,205) ):(131,84) ):(167,181)**

### 3.2. Message Decryption:

The Subgroup Key ($S_k$) is (155,115)
Random no. Chosen can be found by the $K_A G$ Value (32,108)
From (32,108) we may trace the value $K_A$ as 75.

$S_k$ = (155,115) = 229 G
$K_A S_K$ = 75*229G = 64

1) $PM + K_A S_K = (16,100)$
   $PM = (16,100) - K_A S_K = (16,100) - 64 = 164 - 64 = 100\,G \Rightarrow (93,77)$ → **d**

(2) $PM + K_A S_K = (72,197)$
   $PM = (72,197) - K_A S_K = (72,197) - 64 = 168 - 64 = 104\,G \Rightarrow (163,50)$ → **h**

(3) $PM + K_A S_K = (133,163)$
   $PM = (133,163) - K_A S_K = (133,163) - 64 = 103 - 64 = 49\,G \Rightarrow (207,96)$ → **1**

(4) $PM + K_A S_K = (164,197)$
   $PM = (164,197) - K_A S_K (164,197) - 64 = 110 - 64 = 46\,G \Rightarrow (83,124)$ → **.**

(5) $PM + K_A S_K = (12,205)$
   $PM = (12,205) - K_A S_K = (12,205) - 64 = 176 - 64 = 112G \Rightarrow (58,12)$ → **p**

(6) $PM + K_A S_K = (131,84)$
   $PM = (131,84) - K_A S_K = (131,84) - 64 = 174 - 64 = 110G \Rightarrow (164,197)$ → **n**

(7) $PM + K_A S_K = (167,181)$
   $PM = (167,181) - K_A S_K = (167,181) - 64 = 167 - 64 = 103G \Rightarrow (67,154)$ → **g**

Therefore decrypted message is **dh1.png.** The implementation is shown in figure 25.

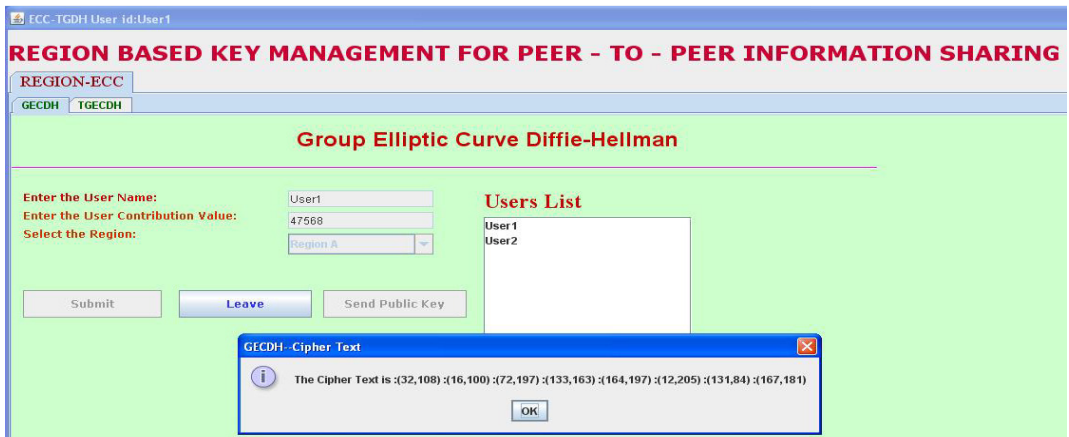

Figure 25. Message Encryption

.

## 4. EXPERIMENTAL RESULTS AND DISCUSSION

The experiments were conducted on a Ad-Hoc System (Laptop) running on a 2.4 GHz Pentium CPU, with 2GB of Memory and 802.11 b/g 108 Mbps Super G PCI wireless cards with Atheros chipset. To test this project in a more realistic environment, the implementation is done by using Java, in an ad hoc network where users can securely share their data. This project integrates with a peer-to-peer (P2P) communication module that is able to communicate and share their messages with other users in the network which is described below.

### 4.1. Searching for information [4, 5, and 6].

  a) Transferring information within the same peer group:
     When the peer group G is in need of searching for any messages within its members ($P_i$), it needs to send a query message to all the members of G (i.e $P_i$). Each peer, $P_j$ in G, that

receives the query message, checks whether any items match the query and responds directly to $P_i$ with a *query response message* that contains the metadata associated with the items that matched the query.

b) Transferring information between peers:
A peer $P_i$ may request a transfer of information from a peer $P_j$, by sending a *transfer request message* to $P_j$. $P_j$, upon receiving this message checks whether it has the information item associated with the request. If $P_j$ has the item then $P_j$ transfers the requested information to $P_i$.

## 4.2. Local Region View

After the establishment of ECRBGKA protocol, the requested peer system files are displayed as shown in the figure.26.

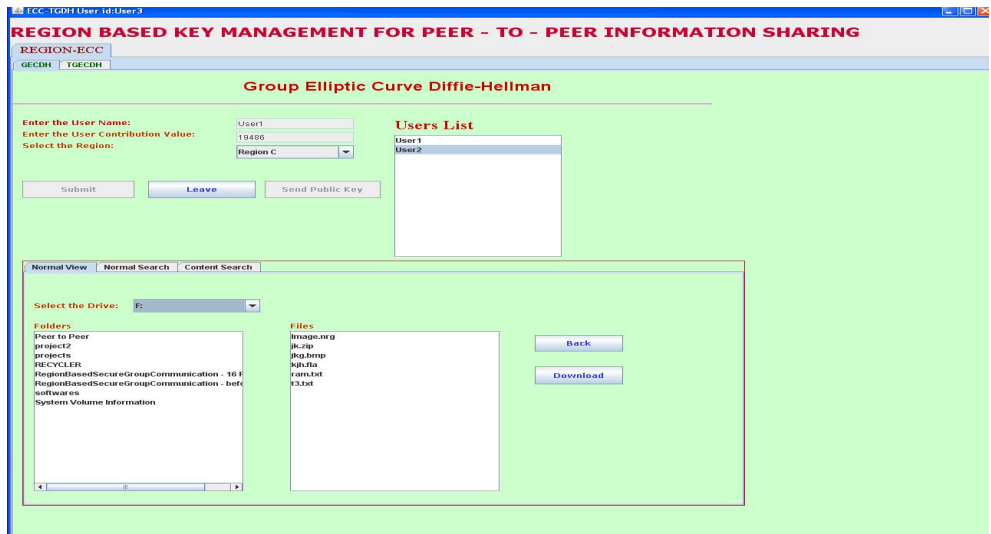

Fig.26.Local Regional view

## 4.3. Normal Search

As soon as the filename is entered and search button gets clicked, the path of the searched file will be enlisted and the desired file can be obtained in a fraction of time. Normal file search is performed as shown in the figure.27.

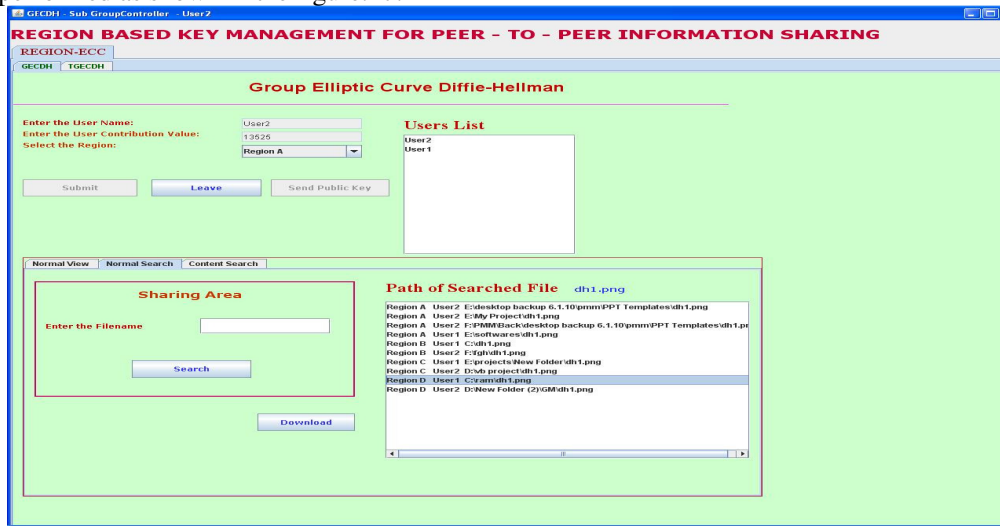

Fig.27. Normal Search

### 4.4. Content Search:

In this kind of search, when we give a content of a file, all files which have that content will be display instantly after clicking over the search button. Content search is performed as shown in the figure 28.

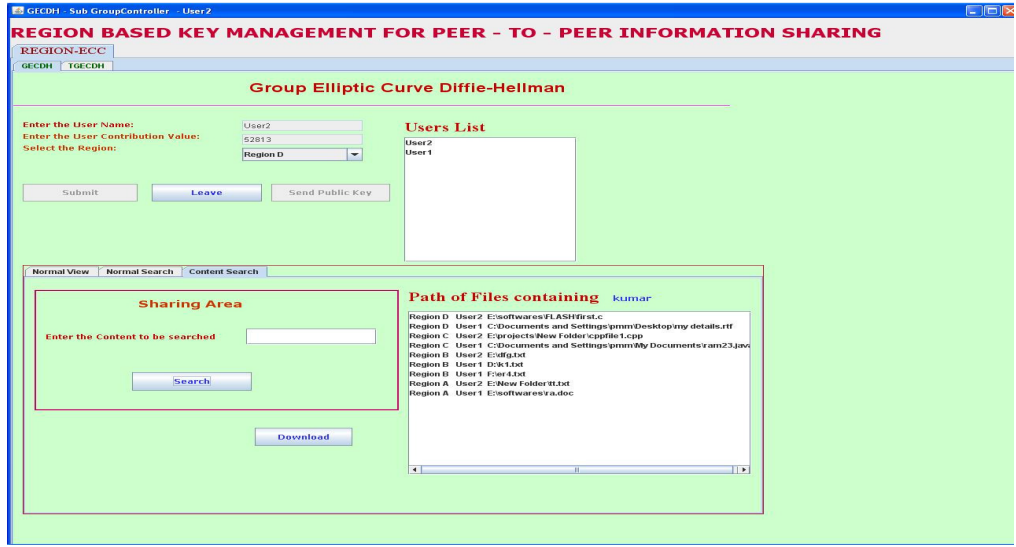

Fig.28. Content Search

## 5. PERFORMANCE ANALYSIS

The performance of the proposed scheme is analyzed in terms of the storage overhead, communication overhead and the computation overhead.

### 5.1. Storage overhead

Storage overhead can be considered as the memory capacity required for maintaining the keys, which is directly proportional to the number of members if the key size are same. In this section, the storage cost is formulated, both at gateway member and at each member. Thus our approach consumes very less memory when compared to TGDH and GDH. TGDH and GDH occupy large memory when members go on increasing. But our Region-based Approach takes very less memory even when the members get increased. Consider (Figure- 29) there are 1024 members in a group our Region-based approach consumes 10% of memory when compared to GDH and 5% when compared to TGDH. The ratio of memory occupied is very less in our approach.

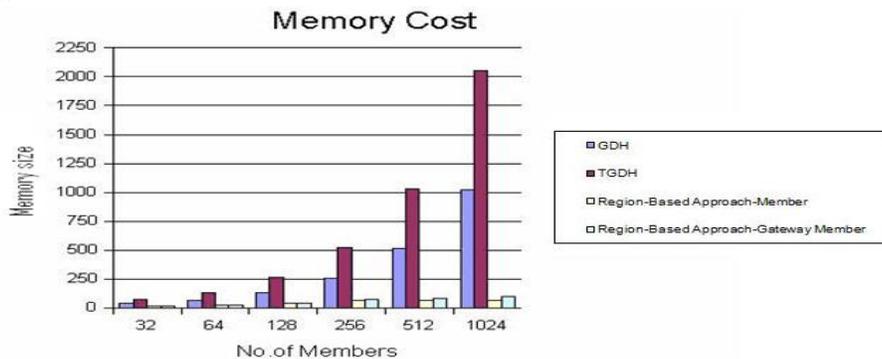

Figure.29. Memory Cost

### 5.2. Communication Overhead:

GDH is the most expensive protocol. TGDH consumes more bandwidth. The Communication and computation of TGDH depends on trees height, balance of key tree, location of joining tree, and leaving nodes. Hence GDH has more communication efficiency than TGDH protocol. But our approach depends on the number of members in the subgroup, number of Group Controllers, and height of tree. So the amount spent on communication is very much less when compared to GDH and TGDH.

Consider (Figure.30&31) there are 512 members in a group our approach consumes only 10% of Bandwidth when compared to GDH and TGDH. So our approach consumes low Bandwidth.

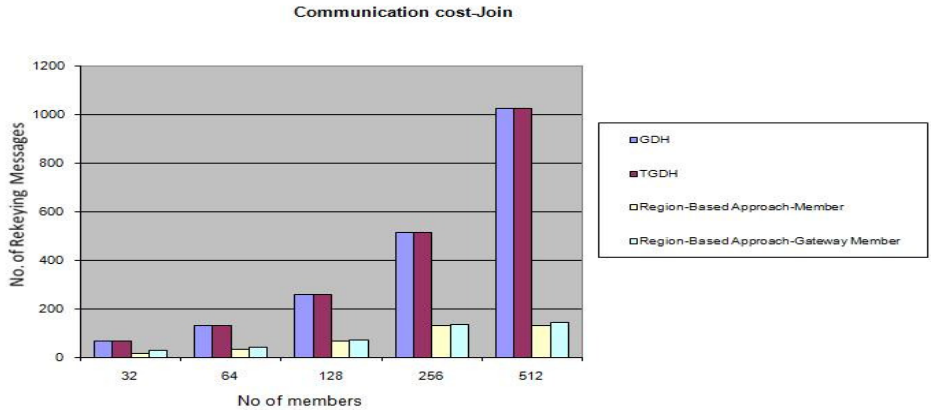

Figure.30. Communication Cost – Join

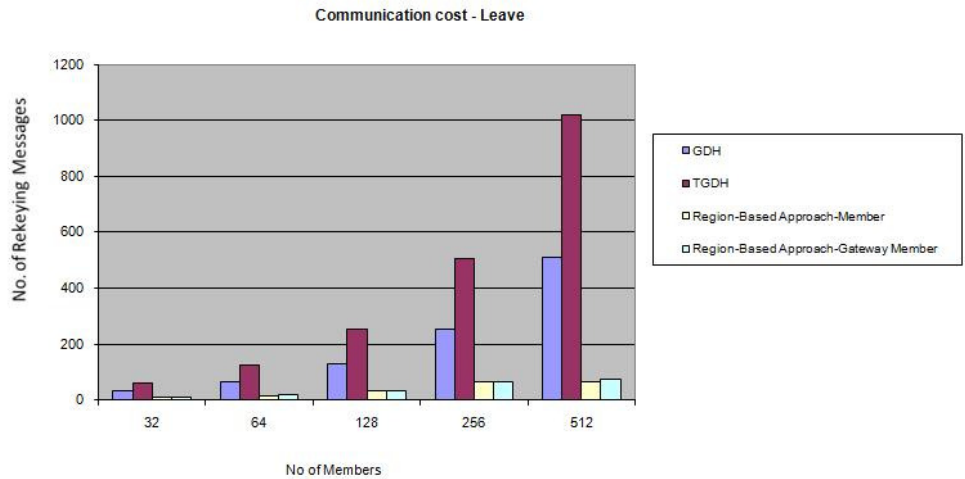

Figure.31. Communication Cost - Leave

### 5.3. Computation Overhead:

The figure.32 shows that Group Elliptic Curve Diffie-Hellman (GECDH)& Tree based Group Elliptic Curve Diffie-Hellman schemes have lower computation time than Group Diffie-Hellman (GDH) schemes for member join operations. The computation time of GDH is **2.2 times** that of GECDH and TGDH is **1.7** times that of TGECDH on average for member join operations. The computation time for member leave operations of TGECDH schemes are far better than group Diffie-Hellman schemes for member leave operations as shown in the

figure.33. Performance wise our approach performs better than GDH&TGDH methods, even for the very large groups.

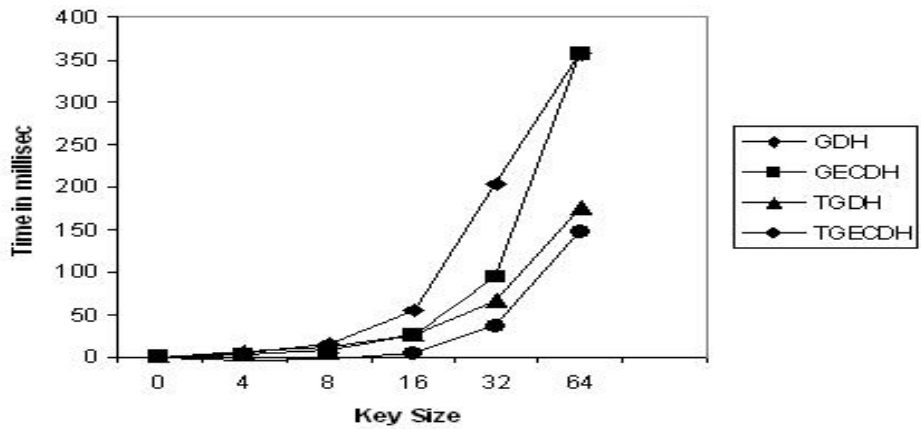

Figure.32. Computation time for Member Join

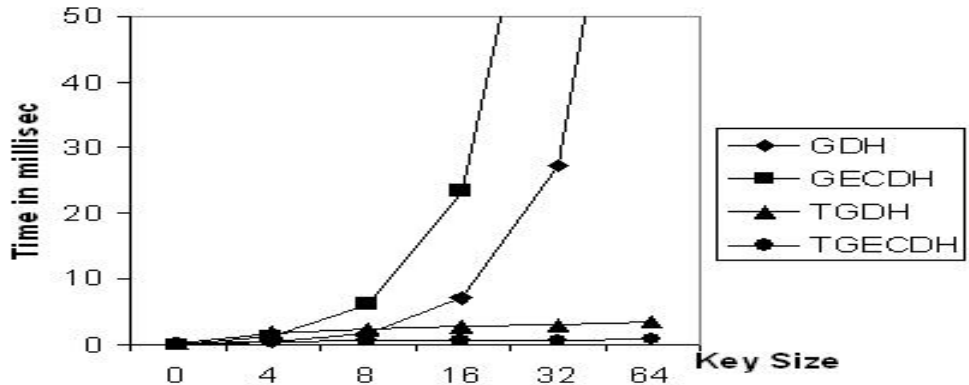

Figure.33 .Computation time for Member Leave

The performance of GECDH&TGECDH over wireless ad hoc Networks can be summarized as follows:
- ➢ It uses smaller keys.
- ➢ It uses less computation time than the DLP-based scheme for the same security level.
- ➢ Smaller packets are used to handle high bit error rate in Wireless links.

## 6. CONCLUSION AND FUTURE WORK

In this paper, we propose and evaluate a new ECDLP-based Diffie-Hellman protocols for secure Peer –to- Peer Information sharing in mobile ad hoc networks. The experiment results show that ECRBGKA scheme is the best protocol in terms of overall performance for secure Peer –to- Peer Information sharing in mobile ad hoc networks. Secure group information retrieval is most efficient if the group members have a common, shared key for securing the communication. Region-Based Group Key Agreement in Peer-to-Peer Information sharing provides efficient, distributed, mechanisms for re-keying in case of dynamic groups, which generates a new key every time a membership change occurs to preserve forward secrecy. We have presented a Region-Based GKA of security mechanisms that can be used to provide ECC based security for Peer-to- Peer information sharing. Our solutions are based on established and proven security techniques and we utilize existing technologies. We make use of these mechanisms to provide efficient, scalable and secure delivery of queries and responses. Our future work will unsolve formalizing these protocols.

**Authors :**

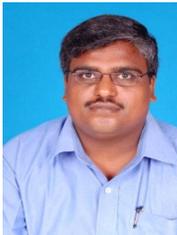

**Kumar. K** received the B.E. Degree in Computer Science and Engineering from the University of Madras, Tamil Nadu, India, in 1999, M.E. Degree in Computer Science and Engineering from Government College of Technology, Coimbatore, Anna University Chennai, India, in 2005. He is an Asst.Professor in the Department of Computer Science and Engineering, Government College of Engineering, Bargur, Tamil Nadu, India. Currently he is working towards Ph.D. in Computer Science and Engineering from the Anna University, Coimbatore. He has presented a number of papers in various National and International conferences. Many of his papers were published in IEEE Explore and Springer. He has to his credit Five International Journal Publications in reputed journals. His research interests include Network Security, Cryptography, Ad Hoc Networks, Distributed System, Peer-to-Peer systems, Wireless Security and Multimedia security.

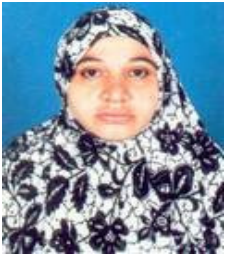
**Nafeesa Begum:** Jeddy Nafeesa Begum is presently working as Assistant Professor (Senior Grade) in Government College of Engineering, Bargur, Krishnagiri District .Tamil Nadu.

She has over 15 years of Teaching Experience. She did her B.E (Computer Science and Engineering) Degree in 1995 from University of Madras and M.E (Computer Science and Engineering) Degree in 2005 from Anna University, Chennai. She is currently pursuing her doctoral programme. Her research interest includes Access Control, Steganography, Network security , Cryptography and Applied Mathematics. She has published in about five international journals and has presented papers in about 5 International Conferences .Two of her papers has been indexed in IEEE Explore.

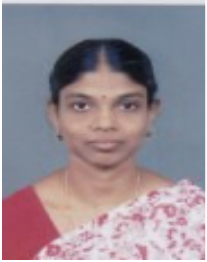
**Dr.V.Sumathy** is presently working as Associate Professor in Government College of Technology, Coimbatore, Tamil Nadu, India. She did her B.E and M.E Degree from Bharathiar University in 1988, 2000. Ph.D Degree in 2007 from

Anna University Chennai. She is presently guiding 10 Research Scholars for their Research Works. She has to her credit many International and National journal Publications .Many of her papers were published in IEEE Explore, Elsevier and Springer. Her research interests include Ad Hoc Networks, Wireless Security and Cryptography.